\documentclass[letterpaper, 10 pt, conference]{ieeeconf}  

\IEEEoverridecommandlockouts                              

\usepackage{graphics} 
\usepackage{graphicx}
\usepackage{ragged2e}
\usepackage{booktabs}
\usepackage{algorithm}
\usepackage{algorithmic}
\usepackage{amsmath} 

\title{\LARGE \bf
Cooperative Pathfinding based on memory-efficient Multi-agent RRT*
}

\author{Jinmingwu Jiang and Kaigui Wu 
}

\makeatletter
\newcommand{\removelatexerror}{\let\@latex@error\@gobble}
\makeatother

\begin{document}

\maketitle
\thispagestyle{empty}
\pagestyle{empty}

\begin{abstract}

In cooperative pathfinding problems, no-conflicts paths that bring several agents 
from their start location to their destination need to be planned. This problem 
can be efficiently solved by Multi-agent RRT*(MA-RRT*) algorithm, which is still 
state-of-the-art in the field of coupled methods. However, the implementation of 
this algorithm is hindered in systems with limited memory because the number of 
nodes in the tree grows indefinitely as the paths get optimized. This paper proposes 
an improved version of MA-RRT*, called Multi-agent RRT* Fixed Node(MA-RRT*FN), which 
limits the number of nodes stored in the tree by removing the weak nodes on the path 
which are not likely to reach the goal. The results show that MA-RRT*FN performs close 
to MA-RRT* in terms of scalability and solution quality while the memory required is 
much lower and fixed.

\end{abstract}

\section{INTRODUCTION}

The problem of planning a series of routes for mobile robots
to destinations and avoiding collisions can be modeled as a 
\textit{cooperative pathfinding} problem. Traditionally,
this problem is often simulated in highly organized environments
such as grids, which include several obstacles and agents.
To find the paths of these agents, the straightforward method is 
looking for the answer in a joint configuration space which is composed
of the state spaces of single agents.
Such a space is typically searched using a heuristic guided function
such as A*\cite{hart_Formal_1968}. However, the problem of cooperative
pathfinding has been proved to be PSAPCE-hard\cite{hopcroft_Complexity_1984}.  

To solve the cooperative pathfinding problems, many works have been proposed 
in the last decades. All these methods can be divided into three categories:
decouple method, coupled method, and hybrid method. Each method has its 
disadvantages, for example, the computational cost of coupled approaches are 
susceptible to the increases of agents, while the decoupled methods cannot 
guarantee their completeness. The hybrid approach, which inherits the advantages 
of the coupled and decoupled approach, seems promising. But when the decoupled 
planner fails, it may be more time-consuming than just using a single planner.

After Karaman and Frazzoli introduce an asymptotically optimal algorithm 
RRT*\cite{karaman_Samplingbased_2011}, \v{C}\'ap marries it to the classical 
multi-agent motion-planning algorithm and proposes Multi-agent RRT*(MA-RRT*)\cite{cap_Multiagent_2013}. 
MA-RRT* is a coupled algorithm. But, unlike other coupled approaches, it alleviates 
the increase of computational cost as the number of agents increases by 
leveraging the idea of the Monte Carlo method. As a result, it can solve the multi-agent path planning problem efficiently. Besides, MA-RRT* comes close to the decouple planner in the efficiency while still maintaining the completeness and optimality, which makes MA-RRT* still advanced in the field of coupled method.

There are many state-of-the-art works that aim to improve the MA-RRT*,
such as \cite{verbari2019multi} in 2019, which improves the efficiency of 
MA-RRT* at the expense of optimality and completeness. Unlike MA-RRT*, \cite{verbari2019multi} 
applies RRT* for each agent in turn, and the agents whose paths have been 
planned by RRT* are treated as moving obstacles.
 
However, the application of the MA-RRT* is hindered in systems with limited memory, 
because as the solution gets optimized, the number of nodes in the tree grows indefinitely.
The closest work to this problem is the RRT* Fixed Nodes(RRT*FN) proposed by Adiyatov\cite{adiyatov_Rapidlyexploring_2013}, 
which only focuses on improving the memory efficiency of RRT*. 
Up to now, none of the works aims to limit the memory required for the MA-RRT* algorithm.

\begin{figure*}[htbp]
	\centering
	\def\svgwidth{2\columnwidth}
	\footnotesize
\begingroup%
  \makeatletter%
  \providecommand\color[2][]{%
    \errmessage{(Inkscape) Color is used for the text in Inkscape, but the package 'color.sty' is not loaded}%
    \renewcommand\color[2][]{}%
  }%
  \providecommand\transparent[1]{%
    \errmessage{(Inkscape) Transparency is used (non-zero) for the text in Inkscape, but the package 'transparent.sty' is not loaded}%
    \renewcommand\transparent[1]{}%
  }%
  \providecommand\rotatebox[2]{#2}%
  \newcommand*\fsize{\dimexpr\f@size pt\relax}%
  \newcommand*\lineheight[1]{\fontsize{\fsize}{#1\fsize}\selectfont}%
  \ifx\svgwidth\undefined%
    \setlength{\unitlength}{1950.75bp}%
    \ifx\svgscale\undefined%
      \relax%
    \else%
      \setlength{\unitlength}{\unitlength * \real{\svgscale}}%
    \fi%
  \else%
    \setlength{\unitlength}{\svgwidth}%
  \fi%
  \global\let\svgwidth\undefined%
  \global\let\svgscale\undefined%
  \makeatother%
  \begin{picture}(1,0.34871203)%
    \lineheight{1}%
    \setlength\tabcolsep{0pt}%
    \put(0,0){\includegraphics[width=\unitlength,page=1]{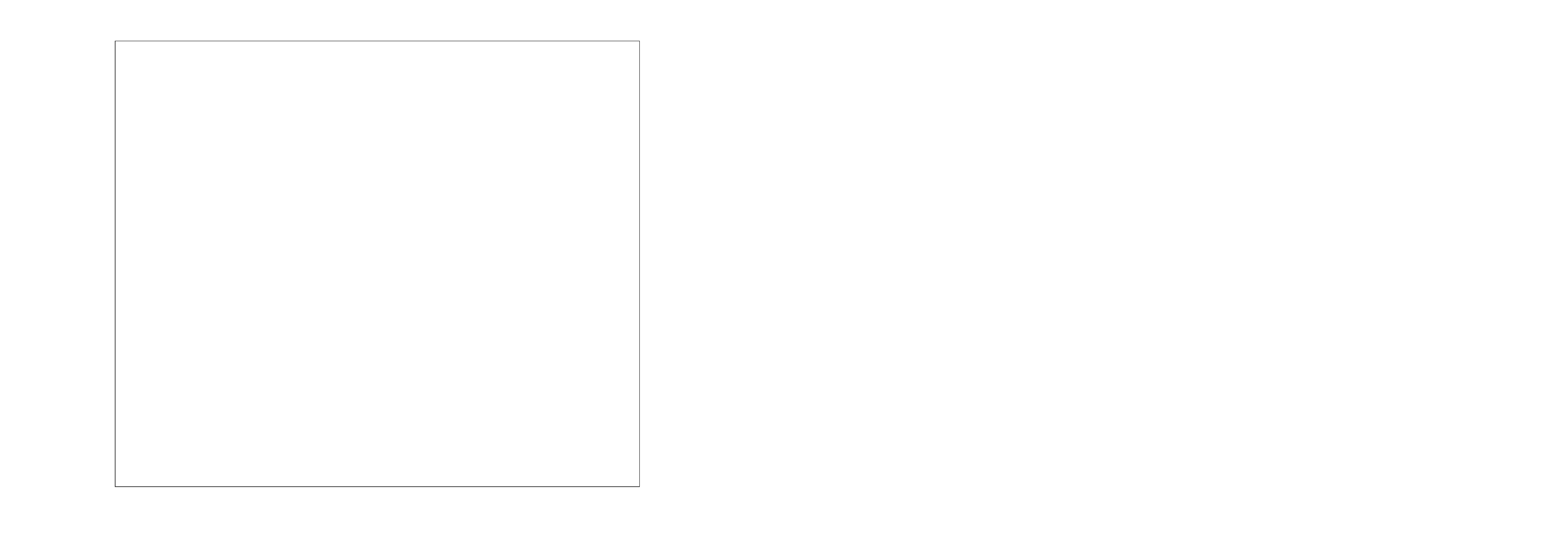}}%
    \put(0.21914671,0.32612457){\makebox(0,0)[lt]{\lineheight{1.25}\smash{\begin{tabular}[t]{l}MA-RRT*\end{tabular}}}}%
    \put(0,0){\includegraphics[width=\unitlength,page=2]{plot_2Dmarrts_marttsfn.pdf}}%
    \put(0.73760161,0.32612457){\makebox(0,0)[lt]{\lineheight{1.25}\smash{\begin{tabular}[t]{l}MA-RRT*FN\end{tabular}}}}%
    \put(0,0){\includegraphics[width=\unitlength,page=3]{plot_2Dmarrts_marttsfn.pdf}}%
  \end{picture}%
\endgroup%

	\caption{single agent navigation using MA-RRT* and MA-RRT*FN respectively.}
	\label{2Dmarrts&marttsfn}
\end{figure*}

This paper presents a new MA-RRT* based algorithm, called Multi-agent 
RRT* Fixed Nodes (MA-RRT*FN), which works by employing a node removal 
procedure to limit the maximum number of nodes in the tree. The property 
of our algorithm can be observed in Fig.\ref{2Dmarrts&marttsfn}, 
which shows the two search trees for single-agent navigation using 
MA-RRT* and MA-RRT*FN respectively in a 2D grid map with 
the same number of iterations. It can be seen that 
the trees MA-RRT*FN generated are more sparse than MA-RRT*.

The main contributions of this paper are as follows: 1) The proposed MA-RRT*FN requires 
a fixed memory, which is much less than MA-RRT* whose memory cost grows indefinitely, 
while its scalability and convergence rate are very close to MA-RRT*. 2) The informed-sampling 
MA-RRT*FN, which is the improved version of MA-RRT*FN, performs very similarly to isMA-RRT* 
concerning the suboptimality of solutions, while its convergence rate and scalability are 
better than isMA-RRT*.

\section{RELATED WORK}
The methods for solving Cooperative pathfinding can be classified 
into three categories: coupled method, decoupled method, and hybrid method.

\textit{1) coupled approaches:} In coupled approaches, all agents' routes are 
computed as a union. The algorithm searches agents' joint configuration space 
to find the solution, which can provide a stronger guarantee on the feasible 
path. For example, Standley\cite{standley_Finding_2010} proposes
two techniques, called Independence Detection(ID) and Operator Decomposition(OD). 
The combination of these two techniques, the ID+OD algorithm, which is capable 
of solving relatively large problems in milliseconds, is both complete and optimal. Standley then refines 
the algorithm into an anytime algorithm called Optimal Anytime(OA)\cite{standley_Complete_2011}, 
which first finds out a solution rapidly, and then utilizes any spare time to 
improve that solution incrementally. 

Alternative methods such as \cite{yu2016optimal} and \cite{erdem2013general} model the Cooperative pathfinding problem as Integer Linear Programming (ILP) and Boolean Satisfiability (SAT) problems respectively. In \cite{yu2016optimal}, Jingjin establishes a one-to-one solution mapping between multi-robot path planning problems and a special type of multiflow network and uses the Integer Linear Programming to optimize the four goals: the makespan, the maximum distance, the total arrival time, and the total distance.

However, All these coupled approaches are susceptible to the increases of agents.
The computation cost of these techniques increases dramatically as the increase 
of robots.

There are also many attempts to use the sampling-based method, such as 
RRT\cite{carpin_Parallel_2002}\cite{lavalle_Rapidlyexploring_1998}\cite{ferguson_Anytime_2006},
PRM\cite{kavraki_Probabilistic_1996}, to solve multi-agent motion planning 
problems. These algorithms alleviate the increase of computational cost as 
the number of agents increases by leveraging the idea of the Monte Carlo 
method. However, for the reason that they are based on the RRT and PRM, these algorithms
are not optimal. After the RRT* is proposed, \v{C}\'ap combines RRT* with OA and 
proposes MA-RRT*\cite{cap_Multiagent_2013}. Unlike other sampling-based method, MA-RRT* is both optimal and complete, and it outperforms many classical algorithms such as ID+OD and OA in terms of efficiency and scalability.

\textit{2) decoupled approaches:} In decoupled approaches, all agents'
paths are planned individually. For example, in \cite{silver_Cooperative_2005},
David Silver introduces three decoupled approaches which decompose the problem 
into several single-agent navigations: Local Repair A*(LRA*), 
Hierarchical Cooperative A*(HCA*) and Windowed Hierarchical 
Cooperative A*(WHCA*). In \cite{geramifard_Biased_2006}\cite{desaraju_Decentralized_2012}\cite{regele_Cooperative_2006}\cite{vandenberg_Prioritized_2005},
the path of each agent is computed individually based on the pre-assigned priorities. 
The same case can also be seen in recent work \cite{verbari2019multi}\cite{ragaglia2016multi}.

Alternative method utilizes conflict based search(CBS)\cite{sharon2015conflict} to find the solution.
Such as the ByPass-CBS and Continuous-Time-CBS proposed by recent work \cite{boyrasky2015don} and \cite{andreychuk2019multi},
which pushes the performance of CBS further.

Although those decoupled methods can efficiently find the solution, their completeness
cannot be guaranteed.  

\textit{3) hybrid approaches:} The hybrid approaches, which leverage the strengths of both
coupled and decoupled techniques, find the solution by firstly employing decouple methods.
If the decoupled techniques fail, the coupled approaches would be employed. For example,
M*\cite{wagner2015subdimensional} solves the multi-agent path planning problem by taking 
the decouple manner first, and when the robots' paths conflict, the conflicting agents are 
merged into a meta-agent and planed the path by a coupled planner. 

The recent works based on CBS also take the idea of hybrid approaches, for example, 
MetaAgent-CBS\cite{sharon2012meta}, and its improved version \cite{boyarski2015icbs}, 
which employs the decoupled techniques first and detects the conflicts, then 
merges the conflicting agents and applies coupled methods.

\section{problem formulation}
To make a fair comparison with the MA-RRT* algorithm, which is simulated on graphs, 
the paper tests both the two algorithms(MA-RRT* and MA-RRT*FN) in a four-connected 
grid world $G_M$ and uses the following definition: Assuming that n agents labeled
$1, ..., n$ are running on a Euclidean space, and each agent, which takes up a single 
cell $x_i(i \in [1,n])$ of the grid world, has a unique start location $s_i$ and 
destination $d_i$. For each timestep, all agents can move to its four neighbor cells 
$x'_i(x'_i \in children(G_M,x_i))$ if it is free or stay on its current location\cite{standley_Complete_2011}.
Besides, the transitions are prohibited in which agents pass through each other.

A cell is free means that it will not be occupied by an agent at the end of the 
timestep and does not include an obstacle. The timesteps that a single agent stays 
on a grid cell are represented as $dur(x_i)$. The total number of timesteps $c$ 
that the agent has taken from its start state $s_i$ to the goal location $d_i$ is 
regarded as the cost of the individual agent's path $path_i$. If all the agents can 
reach their goal without collision, then the sum of each path cost is taken as the 
cost of the final solution, which is the metric of solution quality. Formally,
\begin{displaymath}	
cost(p) = \sum_{i=1}^{n}\sum_{x_i \in p_i}dur(x_i)
\end{displaymath}
Where p stands for an n-tuple of paths $(p_1,...,p_n)$. To simplify the 
representation of nodes in the rapidly random tree, this paper uses $x$ to 
represent the n-tuple of position $(x_1,...,x_n)$. The start positions 
of all agents are given as $s$, which is an n-tuple $(s_1,...,s_n)$. 
Similarly, the n-tuple $(d_1,...,d_n)$ is the destination $d$. Thus, a 
node in the tree can be denoted as an n-tuple joint state, and each state 
stands for the position of a single agent.

\section{MA-RRT* and MA-RRT* fixed nodes}
The multi-agent RRT* algorithm is designed based on RRT* algorithm, which 
can expeditiously find a path from a specific start location to a given target 
region in continuous state space by incrementally building a tree\cite{karaman_Samplingbased_2011}. 
When the first solution is found, the RRT* algorithm will continue to improve 
the solution by sampling new random states in the configuration space, which 
would cause to the discovery of a lower-cost path.

The MA-RRT* inherits all the properties of RRT*. The main difference is that, in continuous configuration 
space, if two nodes are mutually visible, then they can be connected. While in 
the discrete space, two nodes can only be connected if a valid path between the 
two nodes can be found by the heuristic search. Thus, The MA-RRT* more like a 
graph version of RRT*(G-RRT*), unless it searches for the shortest path in a configuration 
space which stands for the joint-state of all agents\cite{cap_Multiagent_2013}. 

The algorithm \ref{MA-RRT*} shows the skeleton of MA-RRT* algorithm. It begins with 
a tree that is rooted at the joint initial state $x_{init}$ and continues to sample the random 
state $x_{rand}$ from free joint configuration space before extending the tree to $x_{rand}$. 
This loop will continue until it is interrupted.

The MA-RRT* Fixed nodes(MA-RRT*FN) utilizes the skeleton of the
MA-RRT* algorithm and extends it with some node removing procedures. Therefore, the MA-RRT*FN behaves like
MA-RRT* before the maximum number of nodes is reached, and after the number of nodes reaches a threshold, it
continues to optimize the tree by removing the weak nodes that are not likely on the path reaching the goal
while adding the new node.

The skeleton of MA-RRT*FN is shown in algorithm \ref{MA-RRT*FN}. Initially, the tree grows before the maximum
number of nodes \textbf{M} is attained, after which the MA-RRT*FN removes a node that has one or no child in
the tree before adding a new node. The MA-RRT* and MA-RRT*FN use the same skeleton of EXTEND and GREEDY procedure, 
shown in algorithm \ref{MA-RRT*FN EXTEND(T,x)} and \ref{MA-RRT*FN GREEDY(G_M,s,d)} respectively.

\begin{figure}[!t]
	\removelatexerror
	\begin{algorithm}[H]
		\caption{MA-RRT*}
		\label{MA-RRT*}
		\begin{algorithmic}[1]
			\STATE $V \leftarrow \{x_{init}\};\ E \leftarrow \emptyset$
			\WHILE{not\ interrupted}
			\STATE $T \leftarrow (V,E);$
			\STATE $x_{rand} \leftarrow  SAMPLE$
			\STATE $(V,E) \leftarrow  EXTEND(T,x_{rand})$
			\ENDWHILE
		\end{algorithmic}
	\end{algorithm}
\end{figure}

\begin{figure}[!t]
	\removelatexerror
	\begin{algorithm}[H]
		\caption{MA-RRT*FN}
		\label{MA-RRT*FN}
		\begin{algorithmic}[1]
			\STATE $V \leftarrow \{x_{init}\};\ E \leftarrow \emptyset;$
			\WHILE{not\ interrupted}
			\IF{$M = NodesInTree(v)$}
			\STATE $(V_{old},E_{old}) \leftarrow (V,E)$
			\ENDIF
			\STATE $T \leftarrow (V,E);$
			\STATE $x_{rand} \leftarrow  SAMPLE;$
			\STATE $(V,E) \leftarrow  EXTEND(T,x_{rand});$
			\IF{$M > NodesInTree(v)$}
			\STATE $(V,E) \leftarrow  ForceRemoval(V,E);$
			\ENDIF
			\IF{$No\ ForceRemovalPerformed()$}
			\STATE $(V,E) \leftarrow RestoreTree();$
			\ENDIF
			\ENDWHILE
		\end{algorithmic}
	\end{algorithm}
\end{figure}

\begin{figure}[!t]
	\removelatexerror
	\begin{algorithm}[H]
		\caption{EXTEND(T,\ x)}
		\label{MA-RRT*FN EXTEND(T,x)}
		\begin{algorithmic}[1]
			\STATE $V' \leftarrow V;\ E' \leftarrow E$
			\STATE $x_{nearest} \leftarrow NEAREST(T,\ x)$
			\STATE $(x_{new},p_{new}) \leftarrow GREEDY(G_M,x_{nearest},x)$
			\IF{$x_{new} \in V$}
			\RETURN $G = (V,E)$
			\ENDIF
			\IF{$p_{new} \ne \emptyset$}
			\STATE $V' \leftarrow V' \cup \{{x_{new}}\}$
			\STATE $x_{min} \leftarrow x_{nearest}$
			\FORALL{$x_{near} \in X_{near}$}
			\STATE $(x',p') \leftarrow GREEDY(G_M,x_{near},x_{new})$
			\IF{$x' = x_{new}$}
			\STATE $c' \gets cost(x_{near})+cost(x_{near},x_{new})$
			\IF{$c' < cost(x_{new})$}
			\STATE $x_{min} \leftarrow x_{near}$
			\ENDIF
			\ENDIF
			\ENDFOR
			\STATE $parent(x_{new}) \leftarrow x_{min}$
			\STATE $E' \leftarrow E' \cup {(x_{min},x_{new})}$
			\FORALL{$x_{near} \in X_{near} \textbackslash \{x_{min}\}$}
			\STATE $(x'',p'') \leftarrow GREEDY(G_M,x_{new},x_{near})$
			\IF{$cost(x_{near}) > cost(x_{new}) + cost(x_{new},x_{near})\ \textbf{and}$\\ $x''=x_{near}$}
			\IF{$onlyChild(parent(x_{near}))\ \textbf{and}$ \\$M = NodesInTree(v)$}
			\STATE $RemoveNode(parent(x_{near}))$
			\ENDIF
			\STATE $parent(x_{near}) \leftarrow x_{new}$
			\STATE $E' \leftarrow E' \cap \{(x_{parent},x_{near})\}$
			\STATE $E' \leftarrow E' \cup  \{(x_{new},x_{near})\}$
			\ENDIF
			\ENDFOR
			\RETURN $G' = (V',E')$
			\ENDIF
		\end{algorithmic}
	\end{algorithm}
\end{figure}

\begin{figure}[!t]
	\removelatexerror
	\begin{algorithm}[H]
		\caption{GREEDY($G_M$,\ \textbf{s},\ \textbf{d})}
		\label{MA-RRT*FN GREEDY(G_M,s,d)}
		\begin{algorithmic}[1]
			\STATE $\textbf{x} \leftarrow \textbf{s};\ c \leftarrow 0;\ \textbf{path} \leftarrow (\emptyset,...,\emptyset)$
			\WHILE{$\textbf{x} \ne \textbf{d}$\ and\ $c \le c_{max}$}
			\STATE $(path_i,...,path_n) \leftarrow \textbf{path}$
			\FORALL{$x_i \in x$}
			\STATE $N \leftarrow children(G_M,x_i)$
			\STATE $x' \leftarrow argmin_{x \in children(G_M,x_i)}h(x_i)$
			\STATE $c \leftarrow c+cost(x_i,x'_i);\ path_i \leftarrow path_i \cup (x_i,x'_i);$
			\STATE $x_i \leftarrow x'_i$
			\ENDFOR
			\IF{$not\ COLLISIONFREE(path_1,...,path_n)$}
			\RETURN \textbf{path}
			\ELSE
			\STATE $\textbf{path} \leftarrow (path_1,...,path_n)$
			\ENDIF
			\ENDWHILE
			\RETURN (\textbf{x},\textbf{path})
		\end{algorithmic}
	\end{algorithm}
\end{figure}

Like the MA-RRT*, in each iteration of MA-RRT*FN, the SAMPLE routine randomly chooses a free state in the joint space.
Then, the EXTEND function generates a new node $x_{new}$ in the free space by steering from the nearest node
to the new randomly sample, and then check whether $x_{new}$ is contained in this tree. If so, $x_{new}$
will be deleted from the tree, and the EXTEND function will restart, if not, $x_{new}$ will be added to
the tree. After that, the algorithm searches the nodes that near the $x_{new}$ to construct the nodes
set $X_{near}$ and chooses a node as the parent of $x_{new}$, which makes $x_{new}$ has the lowest cost to
initial state, from $X_{near}$ and $x_{nearest}$. Finally, it updates the cost of $X_{near}$ by rewiring to
$x_{new}$ if these nodes decrease the total cost by assigning $x_{new}$ as the parent. 

Unlike MA-RRT*, the MA-RRT*FN employs a node removing procedure in the EXTEND function,  shown on lines 24 
and 25. During the EXTEND procedure, the algorithm updates the cost
of nodes near the newly added node $x_{new}$. If a node
$x_{near}$ from $X_{near}$ could reach a lower cost to the initial state by reconnecting to the newly added
node, then the algorithm would check whether the parent of this node has only one child and whether the number
of nodes in the tree reaches M. If so, $x_{near}$ will be rewired as a child of $x_{new}$, and the parent of
$x_{near}$ will be deleted. If none of the nodes in the near domain of $x_{new}$ has only one child to remove,
then the \textit{ForcedRemoval} procedure in algorithm \ref{MA-RRT*FN} will be employed, which searches the 
entire tree, except the $x_{new}$ and the goal node, to find the nodes without children and deletes one
randomly\cite{adiyatov_Rapidlyexploring_2013}. In case no nodes are deleted in \textit{EXTEND} and
\textit{ForceRemoval} function, $x_{new}$ is removed from the tree.

MA-RRT*FN has the same GREEDY procedure as MA-RRT*. In the GREEDY procedure, the joint state is decomposed 
to n single-agent states. Thus, the algorithm can steer
each agent from its start node \textbf{s} to the destination \textbf{d} for one timestep separately by merely
depending on heuristic guided search, which utilizes Euclidean distance as the metric, and then check the path
generated for all agents collide or not. If those paths are conflicted, the algorithm will return the path
calculated in the prior timestep; if not, the algorithm will check whether all agents reach the target, if
they do, the algorithm would return the path of all agents as a series of joint transitional states between
the \textbf{s} and \textbf{d}, forming an edge in the tree. If the goal is not attained and the cost of paths
exceeds the user-specified threshold $c_{max}$, the algorithm will return the path between the \textbf{s} and
the currently arrived node.

\begin{figure}[!t]
	\removelatexerror
	\begin{algorithm}[H]
		\caption{isMA-RRT*FN}
		\label{isMA-RRT*FN}
		\begin{algorithmic}[1]
			\WHILE{not\ interrupted}
			\FOR{$i = 1...n$}
			\STATE run the G-RRT* algorithm for agent i
			\ENDFOR
			\IF{all agnents find the paths though G-RRT*}
			\STATE run MA-RRT*FN algorithm based on biased sampling
			\ENDIF
			\ENDWHILE
		\end{algorithmic}
	\end{algorithm}
\end{figure}


Both MA-RRT* and MA-RRT*FN evenly sample the random states in agents’ joint configuration space, which would
cause a relatively lower convergence rate. To improve the speed of MA-RRT*FN in finding the solutions, we take
the ideas from isMA-RRT*, the improved version of MA-RRT* proposed in \cite{cap_Multiagent_2013}. The improved algorithm is 
called informed sampling MA-RRT*FN(isMA-RRT*FN), shown in algorithm \ref{isMA-RRT*FN}, which runs G-RRT* for every single agent 
to find some high-quality solutions and then runs MA-RRT*FN for all agents together with biased sampling, which samples states 
near the single-agent optimal path. 

\section{experiments and results}
The paper first compares the capability of the MA-RRT*, MA-RRT*FN, isMA-RRT* and isMA-RRT*FN in terms
of scalability and suboptimality, then compares the memory cost and solution quality of these algorithms in a
50x50 grid with 3 agents navigation. In the sampling procedure, all four algorithms choose the final goal
state as the new random sample with the probability of \textbf{p}, which is the user-specified parameter, to
speed the procedure of spanning towards the target. All experiments are performed on \textit{matlab 2018a 64-bit} in a common program framework and tested on \textit{intel core i7 8700k} 3.7 GHz CPU.

To make a fair comparison between these four algorithms, this paper utilizes the problem instance set of \cite{cap_Multiagent_2013}, mentioned as follows, to evaluate the capability of the algorithms. The agents run in a grid-like square-shaped world, where each agent occupies a single cell. At each timestep, all agents can stay on the cell waiting for other agents or move to the 4-neighborhood cell of its location if these cells are free. The ten percent of the grids are removed to represent obstacles or barriers. A unique
start location and destination are selected randomly for every agent. 

The problem instances set varies in the following two parameters: The grid sizes: 10x10, 30x30, 50x50, 70x70,
90x90 and the numbers of agents: 1, 2, 3, 4, 5, 6, 7, 8, 9, 10, which are the same to
\cite{cap_Multiagent_2013}. The two parameters are combined in each grid size and number of agents. For each
combination, this paper randomly sets 120 instances. Therefore, the first experiment contains 6000 different
problem instances in total. All algorithms are implemented on the same instance set, and the runtime of each
instance is limited to 5 seconds. For MA-RRT*FN and isMA-RRT*FN algorithm, the maximum number of nodes is set to 200. 

\begin{figure}[tbp]
	\centering
	\scriptsize
	\def\svgwidth{0.9\columnwidth}
	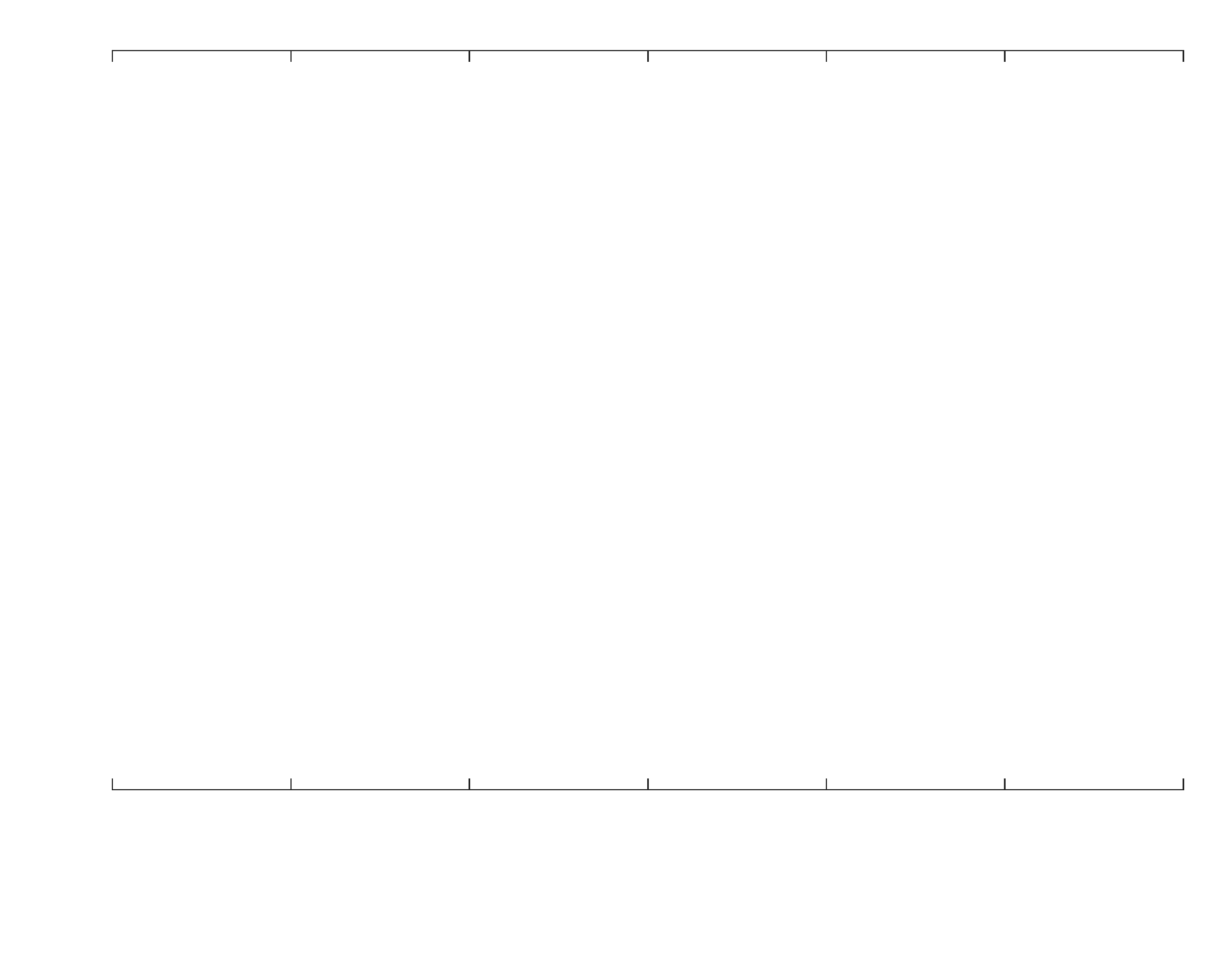
	\caption{Performance curve.}
	\label{Performance_curve}
\end{figure}

\begin{figure}[tbp]
	\centering
	\footnotesize
	\def\svgwidth{0.9\columnwidth}
	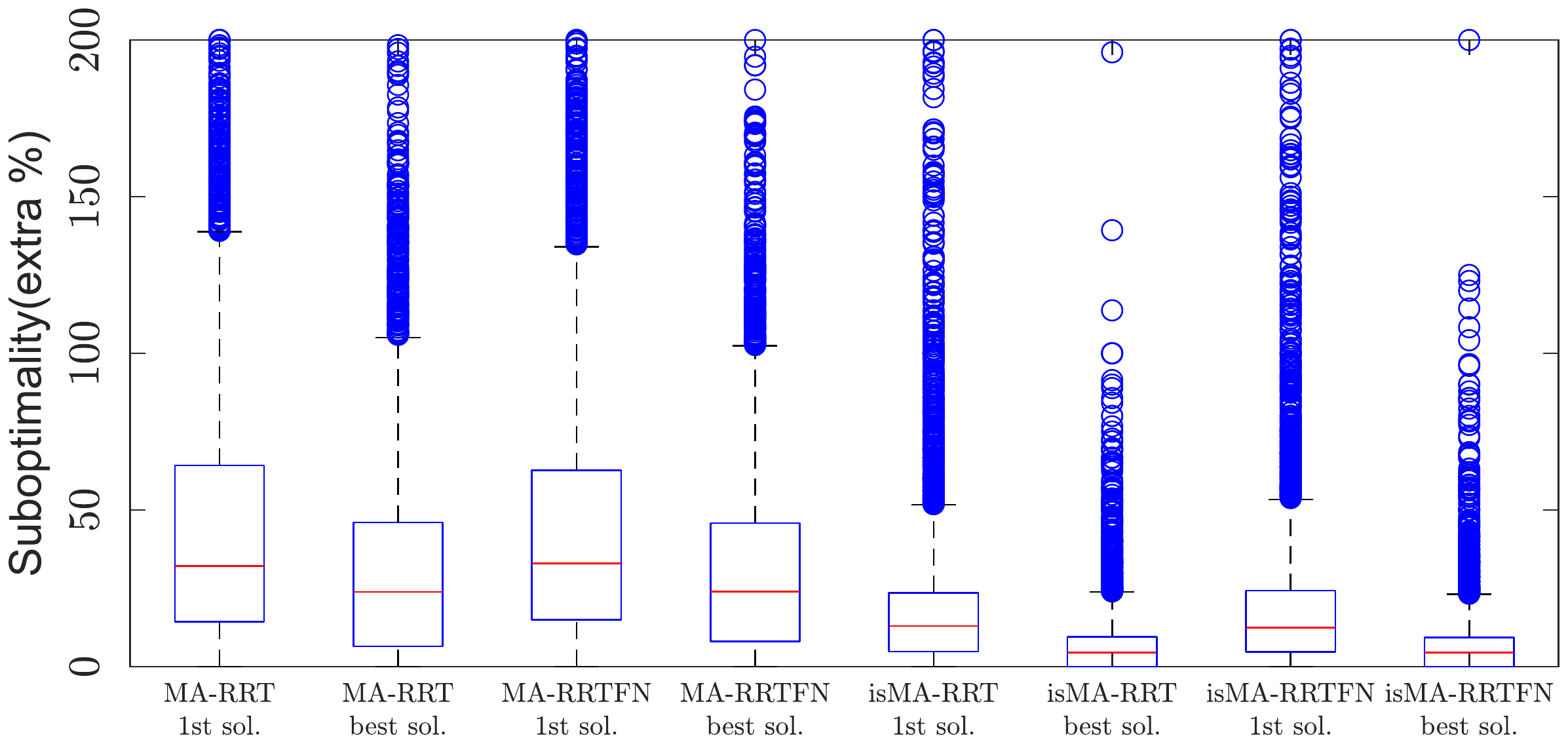
	\caption{Suboptimality.}
	\label{Solution_quality}
\end{figure}

\begin{figure}[htbp]
	\centering
	\scriptsize
	\def\svgwidth{0.9\columnwidth}
	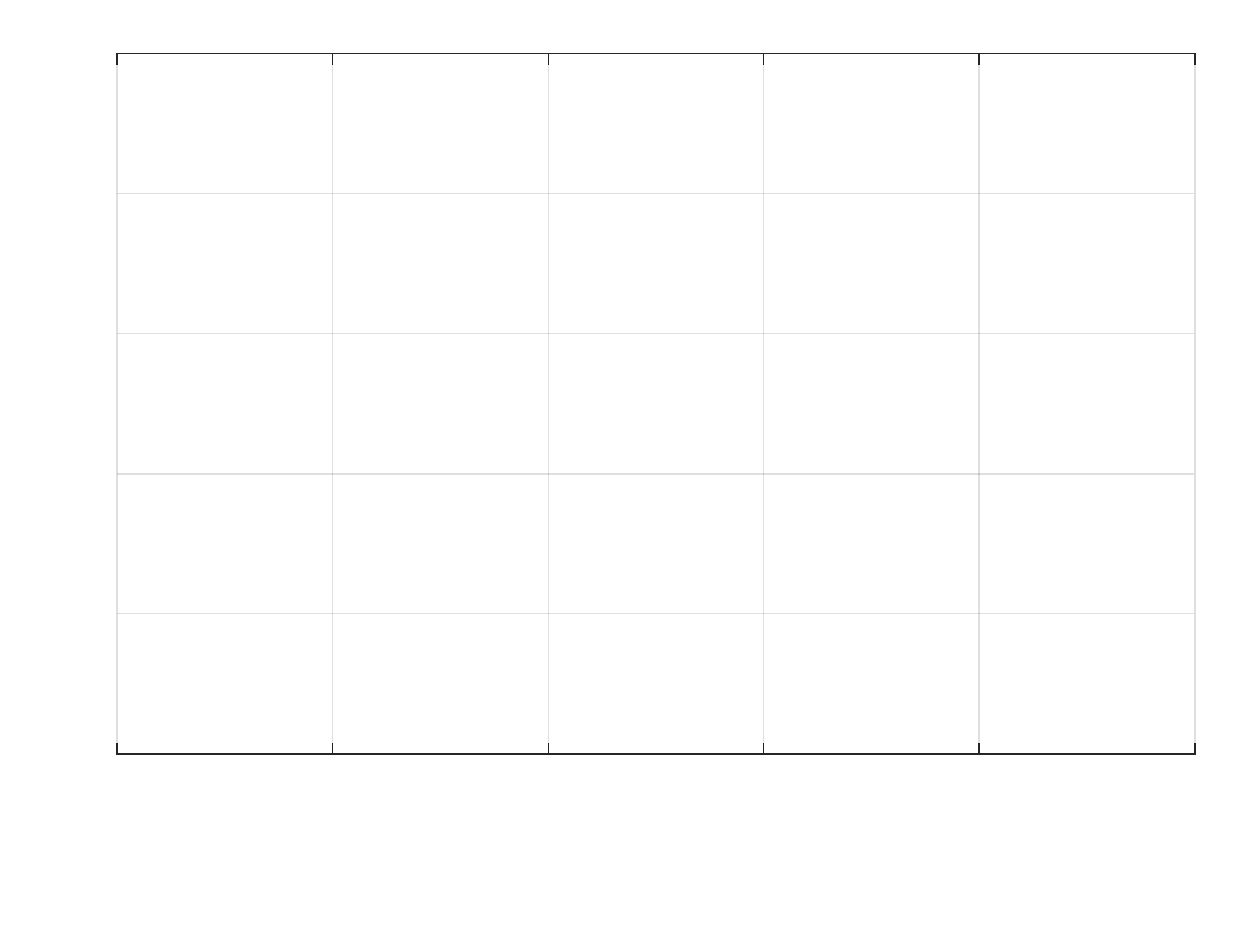
	\caption{Solution quality.}
	\label{Three_Agents_Navigation}
\end{figure}

\begin{figure}[htbp]
	\centering
	\scriptsize
	\def\svgwidth{0.9\columnwidth}
	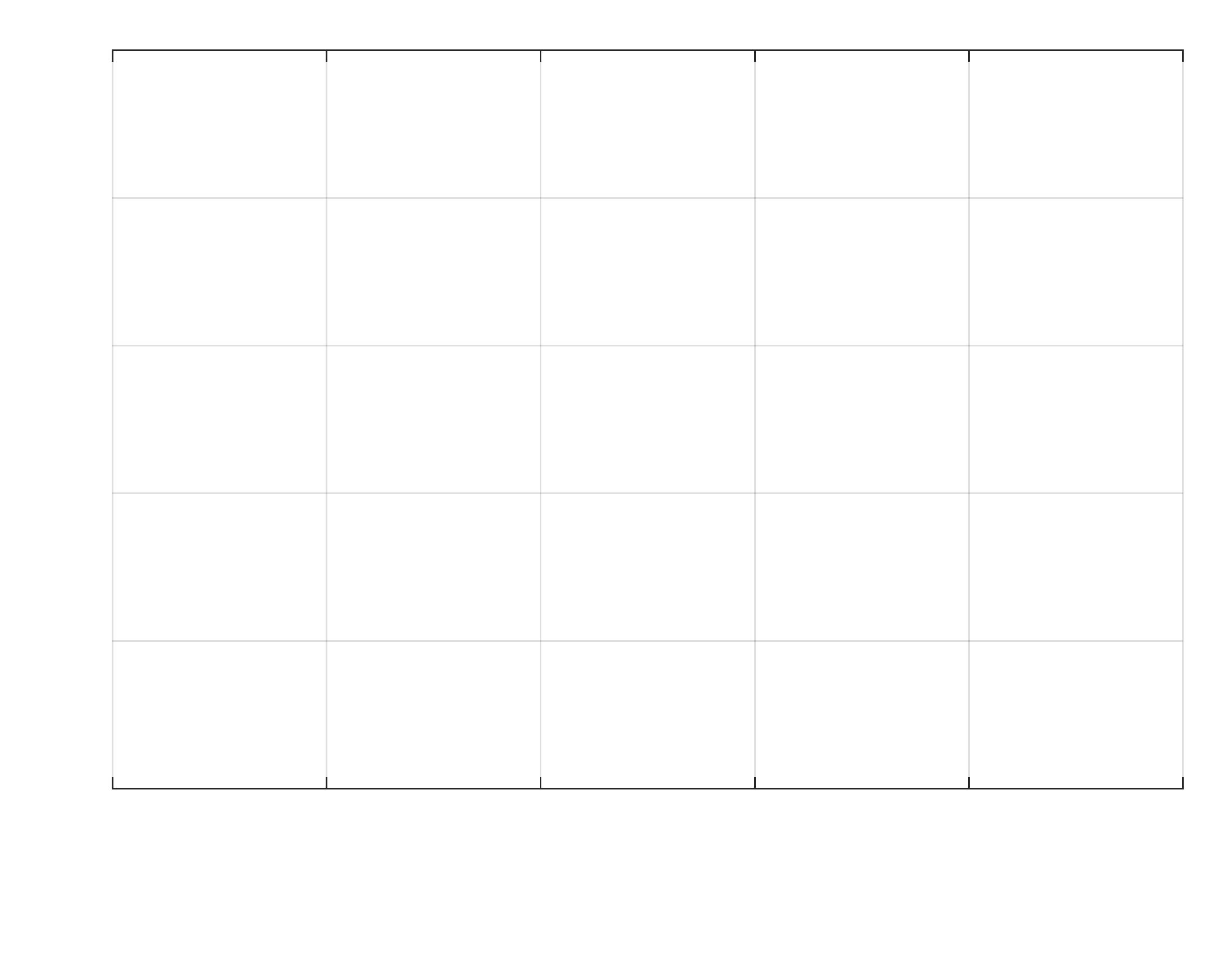
	\caption{Memory required.}
	\label{Memory_required}
\end{figure} 

The results are plotted in Fig.\ref{Performance_curve} and Fig.\ref{Solution_quality}. In Fig.\ref{Performance_curve}, the values in the x-axis are the index of instances which are sorted according to the
runtime needed when the first valid solution is found, the values in the y-axis are the runtime when the
algorithm finds the first solution. For each algorithm, the ordering can be different. The last point of
x-position in the performance curve indicates how many instances are solved within 5 seconds. It can be seen
that MA-RRT* resolves 66\% of the instances, MA-RRT*FN 65\%, isMA-RRT* 86\% and isMA-RRT*FN 87\%, from the
problem instance set. The relative solution quality is shown in Fig.\ref{Solution_quality}. The experiment
compares all algorithms in terms of the first returned solution and the best returned solution within 5 seconds
runtime limit. The suboptimality is calculated by the following formula:

\begin{displaymath}	
	\text{suboptimality} = \left(\frac{\text{the cost of returned solution}}{\text{the\hspace{0.1em} cost\hspace{0.1em} of\hspace{0.1em} optimal\hspace{0.1em} solution}}\ -\ 1 \right) \cdot\ 100.	
\end{displaymath}

As shown in Fig.\ref{Solution_quality}, MA-RRT*FN and isMA-RRT*FN have a similar suboptimality to MA-RRT* and isMA-RRT*FN, respectively.

Then the paper compares the four algorithms in terms of memory cost and convergence rate. For clarity, this experiment fixes the two parameters, the grid sizes:
50x50 and the numbers of agents: 3, to qualitatively show the memory needed and convergence rate of all algorithms. And for this problem instances set, this
experiment randomly sets 120 instances with different random obstacles and different start locations and destinations. All algorithms run on the same instance 
set, and the maximum number of iterations of each instance is limited to 5000. For MA-RRT*FN and isMA-RRT*FN , the maximum number of nodes is set to 1000.

Fig.\ref{Three_Agents_Navigation} and Fig.\ref{Memory_required} show the average minimum path cost and
the average number of nodes in the tree versus the iterations of all algorithms in terms of the solutions of
120 instances, respectively. The x-position of the first point in the Path cost curve can be interpreted as the
solution the algorithm found at the first iteration. For those who do not find a path at the current
iteration, the cost of their first solution will be taken into account to compute the average minimum path
cost at the current iteration.

Fig.\ref{Three_Agents_Navigation} shows that the MA-RRT*FN has a similar convergence rate to MA-RRT* while
its number of nodes in the tree is much less, as shown in Fig.\ref{Memory_required}, memory required for
MA-RRT* grows linearly with the iterations increase, while the number of nodes stored in MA-RRT*FN is lower
and fixed. The results also indicate that the isMA-RRT*FN performs better than isMA-RRT* concerning the
convergence rate to the optimal path, while it also has a lower and fixed memory. Finally, MA-RRT* is
proved to be convergent in \cite{cap_Multiagent_2013}, although the experiment results strongly imply that
the MA-RRT*FN and isMA-RRT*FN also have the theoretical guarantee of converging to the optimal path, the
optimality of MA-RRT*FN and isMA-RRT*FN remains to be proved.

\section{conclusion and future work}
The paper proposes MA-RRT*FN, an anytime algorithm that has lower demands in the memory requirements, to slove the multi-agent path planning problem in the systems with limited storage. Unlike MA-RRT*, whose memory cost is indefinite as the solution converges to the optimal path, our techniques employ some node removing procedures to limit the number of the nodes storing in the tree and keep on optimizing the path when finding the solution in agents' joint-state space. The experiment results show that the MA-RRT*FN, which has a fixed number of nodes in the tree, performs as well as MA-RRT* in terms of scalability, solution quality and convergence rate in solving multi-agent path planning problems. While the improved version, isMA-RRT*FN, has a better convergence rate and scalability than isMA-RRT* while its memory required is much lower.

This paper simulates the algorithm on a motion graph, which connectes the states in the tree by a valid path. However, the algorithm can also be extended to continuous space by using the straight-line visibility approach in place of the GREEDY function.

In the future, we will continue to improve the convergence rate of MA-RRT*FN by employing different node removing procedures. Another area we would like to explore is the application of the MA-RRT*FN algorithm in a more dense environment. 

\addtolength{\textheight}{-12cm}   


\bibliographystyle{IEEEtran}
\bibliography{root}

\end{document}